\documentclass{article}
\usepackage[
    top=3cm,
    bottom=3cm
]{geometry}
\usepackage{graphicx}
\usepackage{subcaption}
\usepackage{xcolor}
\usepackage{placeins}
\usepackage{url}

\title{Comment on the preprint:\\ 
\textit{First Limits on Axion Dark Matter from a DALI Prototype} \\ arXiv:2603.21951
}

\author{
E. Garutti$^{1,*}$,
A. Ivanov$^{2}$,
B. Majorovits$^{2,\dagger}$ \\[1ex]
{\small $^{1}$Universität Hamburg, Hamburg, Germany} \\
{\small $^{2}$MPI für Physik, Garching, Germany} \\
[1ex]
{\small $^{\dagger}$Spokesperson of the MADMAX collaboration} \\
{\small $^{*}$Chair of the MADMAX collaboration board} \\
{\small Corresponding authors: erika.garutti@desy.de, bela.majorovits@mpp.mpg.de}
}

\begin{document}

\maketitle

\begin{abstract}

We report on an independent assessment of the sensitivity of the DALI setup to axion-like particle (ALP) dark matter, as described in \cite{demiguel2026limitsaxiondarkmatter}. Our analysys is based on the information provided in the preprint and using the formalism developed within the MADMAX collaboration. We do not identify any mode below 12 GHz with significant coupling to the axion field to account for the reported sensitivity. We provide recommendations for assessing the axion coupling of electromagnetic modes in dielectric haloscope experiments.

\end{abstract}

\clearpage

\begin{figure}[h]	
	\centering	
	\includegraphics[width=1\textwidth]{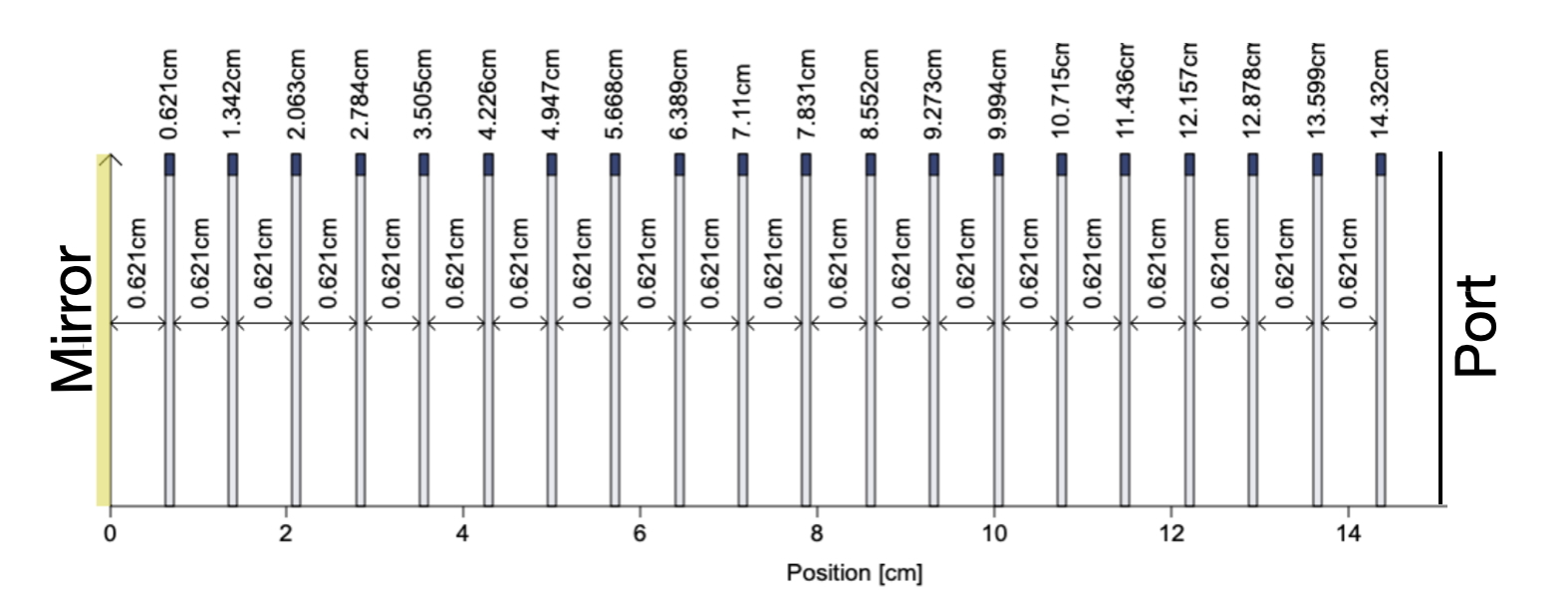}
	\caption{Details of our simulation based on the description from DALI's pilot run. The exact coordinates of the 20 plates are shown for the reported fixed spacing of 6.21 mm. The resonator is enclosed by a perfect lossless mirror and a perfect antenna (implemented via the port boundary). The material parameters of the plates are according to the nominal reported: relative dielectric constant $\epsilon_r = 30$, loss tangent tan$\delta$ = 10$^{-4}$ and thickness of 1 mm.
	}	
	\label{fig:dali_stack}
\end{figure}

In this comment we outline our attempt to reproduce the results reported by the DALI collaboration in the preprint  \cite{demiguel2026limitsaxiondarkmatter}. Particularly, we investigate the ability of the discussed apparatus to couple to the axion-induced field over the broad range of 5 to 7.5 GHz. To do this, we explicitly simulated the resonator stack as described in the \textit {Experimental setup} section of the preprint   \cite{demiguel2026limitsaxiondarkmatter}.
The geometrical details are shown in Figure \ref{fig:dali_stack}.

We implement the reported geometry in COMSOL assuming an idealized setup: the electromagnetic field is perfectly plane-wave-like, no edge scattering and no transverse higher order mode conversion. The electric field is polarized tangentially to the plate surface. As a result, the simulation implements the idealized case of unbound infinite  disks, where the field distribution is controlled solely by the positions of the disks and mirror as well as the dielectric constant $\epsilon_r$.

\begin{figure}[]	
	\centering	
	\includegraphics[width=0.8\textwidth]{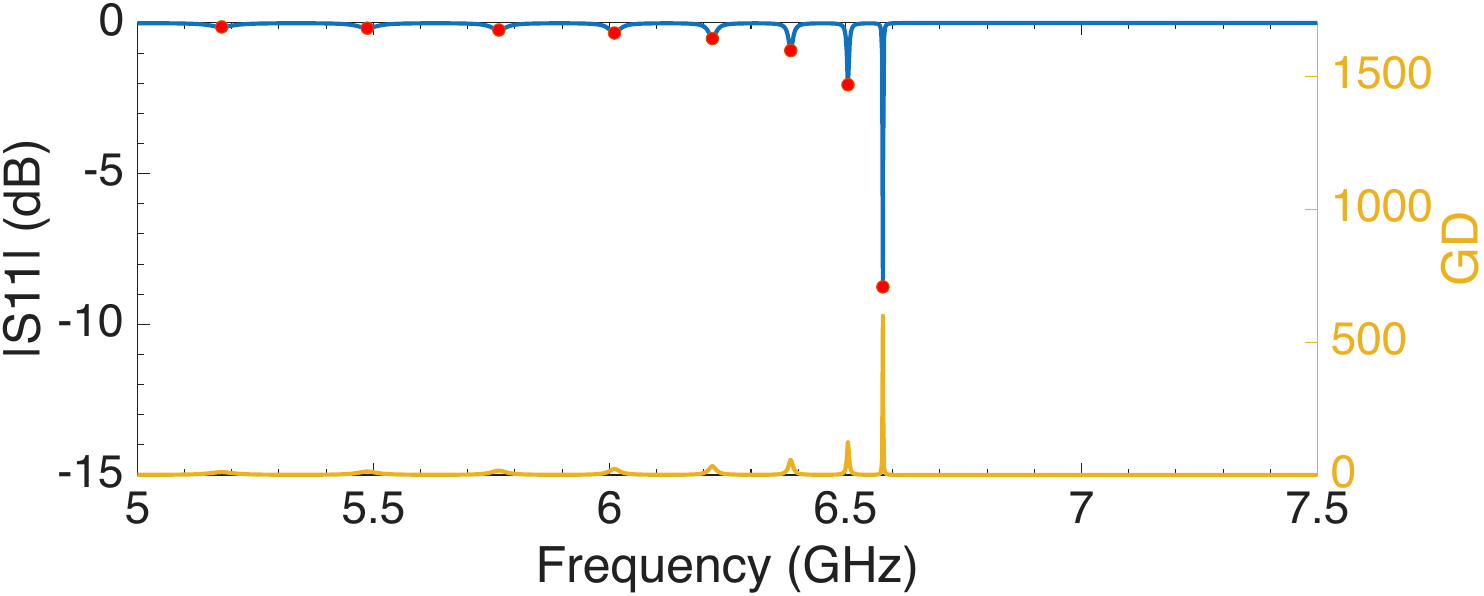}
	\caption{Simulated electromagnetic response based on the description from DALI's pilot run representing idealized setup with  infinitely large disks and no higher-order modes. The reflectivity (in blue) and group delay GD (in yellow) are shown. The resonance frequencies are indicated by the red dots. }	
	\label{fig:S11}
\end{figure}

\begin{figure}[]	
	\centering	
	\includegraphics[width=1\textwidth]{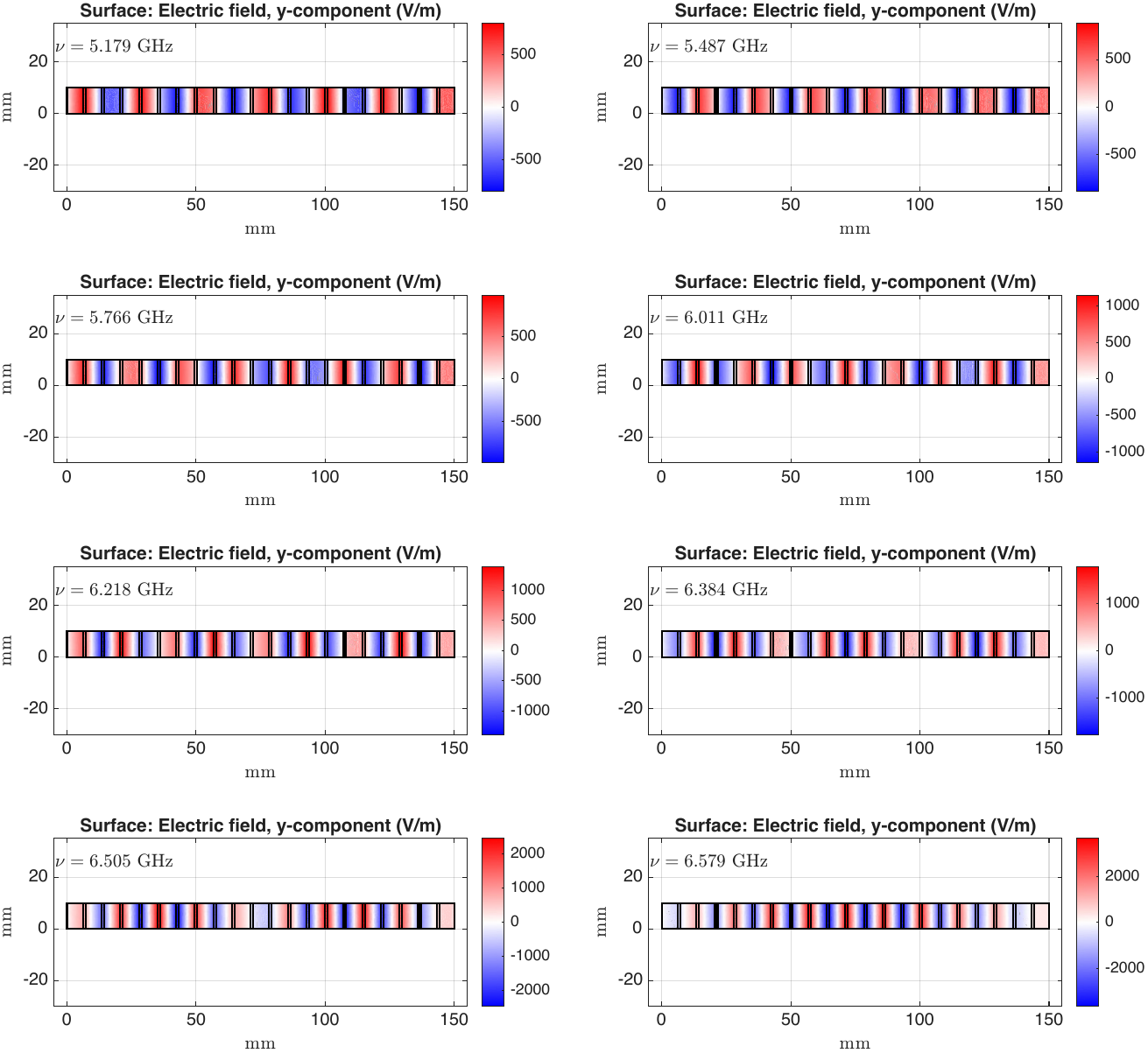}
	\caption{On-resonance electric field distribution maps at the indicated resonant frequencies, obtained directly from the COMSOL solution. The electric field is polarized along the $y$-direction, tangential to the dielectric plates, and varies only along the longitudinal propagation direction ($x$). The $y$-dimension is only for visualization, as the solution is independent on the transverse boundaries, consistent with the assumption of infinite plates.}	
	\label{fig:E_field}
\end{figure}

The simulated reflection response is evaluated at the port and it is shown in Fig. \ref{fig:S11}. From  the group delay GD and the magnitude of the reflection coefficient S11 we clearly identify eight distinct resonances which are summarized in Table \ref{tab:resonances}. In addition, in Fig.\,\ref{fig:E_field}, we include the on-resonance electric field distribution evaluated at the eight resonance frequencies.  

\newpage
To evaluate the power coupling of the resonant modes to the axion-induced field we use the well-known normalized form factor \cite{Cavity_design_RSI_2015}:
\begin{equation}
	C=
	\frac{
		\left|\int_{V} dV\, \mathbf{B}_e \cdot \mathbf{E}_m \right|^2
	}{
	 V \left|\mathbf{B}_e\right|^2
		\int_{V} dV\, \left|\mathbf{E}_m\right|^2
	},
\end{equation}
where $\mathbf{B}_e$ is the external magnetic field considered homogeneous,   $\mathbf{E}_m $ is the standing wave electric-field eigenfunction of the resonance and the integration is along the volume $V$ of the resonator. The resonances are independent of the transverse boundary, hence the form factor is separable to longitudinal and transverse parts. For the assumed perfect antenna the transverse coupling factor is unity so the power coupling depends only on the amount of overlap of the electric field to the constant $\mathbf{B}_e$ along the longitudinal coordinate.

Next we perform this calculation numerically, from the field distributions shown in Figure \ref{fig:E_field}. The results for the numerically evaluated form factors are included in Table\,\ref{tab:resonances}. It is seen, that for all the resonances in the studied frequency range the form factor is marginal and hence the power coupling is negligible. 

\begin{figure}[h]
	\centering
	\includegraphics[width=0.9\linewidth]{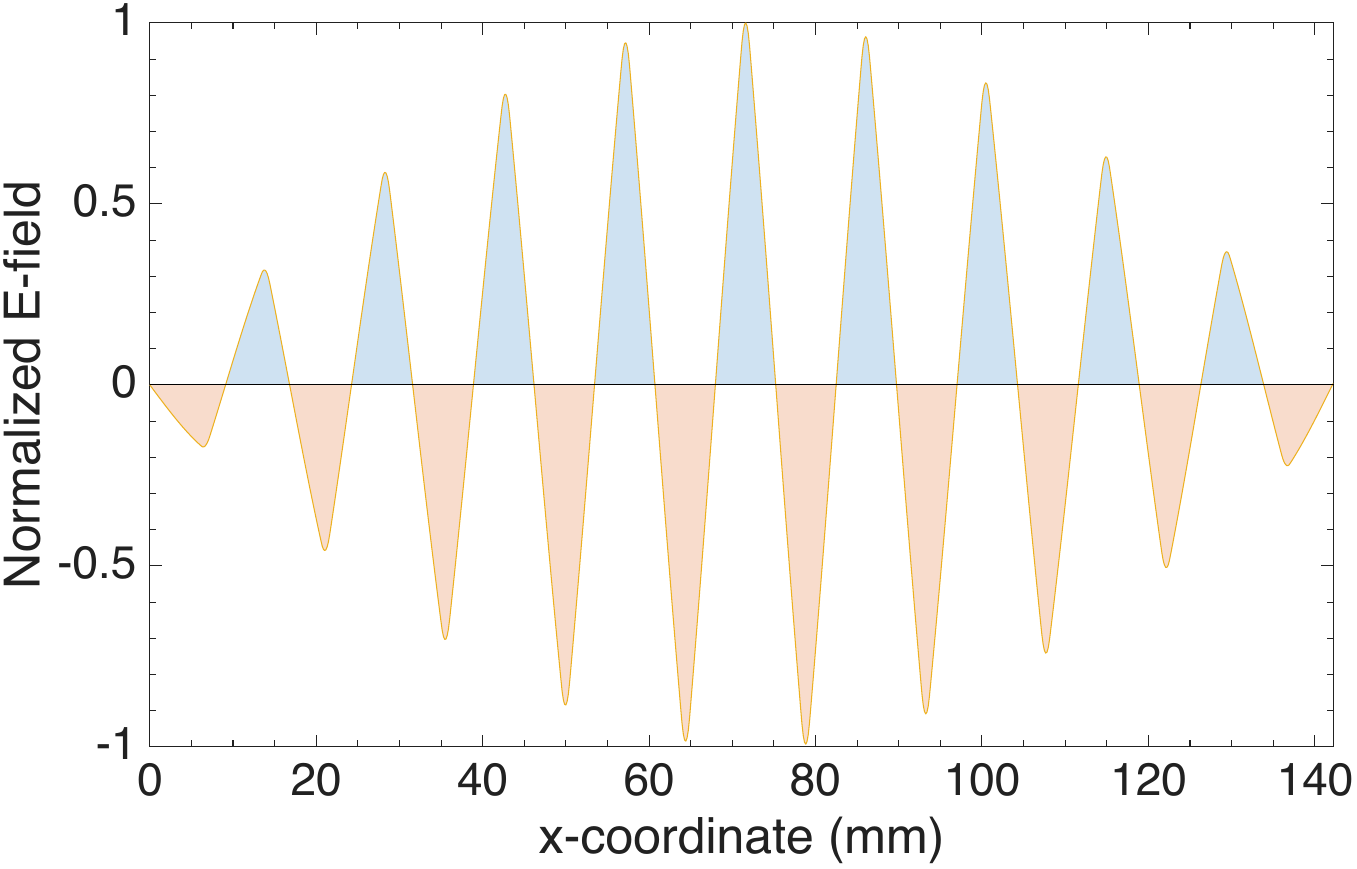}	
	\caption{Longitudinal distribution of the electric field inside the resonator stack shown for the resonance at $\approx 6.58$ GHz.}
	\label{fig:esnapshot}
\end{figure}

To illustrate why, in Figure \ref{fig:esnapshot}, we emphasize the field distribution of the resonance at $\approx\,6.58$\,GHz as it is closest to the reported 6.907\,GHz\footnote{The resonance reported in the preprint is found at ~6.9 GHz, whereas the closest resonance predicted by our simulation is found at  $\approx$6.58 GHz. Such discrepancies are very common, as small variations in the dielectric constant and geometric tolerances can readily shift the resonance frequencies by several hundred megahertz. For this reason, our analysis covers a broad frequency range around the reported resonance and examined the complete mode spectrum between 5 and 12 GHz.}. It is seen that the positive and negative lobes practically cancel out, leading to negligible overlap which explains the numerical result reported in the table. We have also examined the spectrum in the range 7.5--12 GHz, but no resonances are present, which is expected behavior given the periodic nature of the geometry.

\begin{table}[htbp]
	
	\centering
	
	\caption{Resonance frequencies, reflectivity, and form factors obtained from the simulation for the eight resonances within the investigated frequency range.}
	
	\label{tab:resonances}
	
	\begin{tabular}{ccc}
		
		\hline
		
		Frequency (GHz) & $S_{11}$ (dB) & Form factor C \\
		
		\hline
		
		5.1788 & -0.1343 & 0.0028 \\
		
		5.4872 & -0.1745 & $1.1212\times10^{-4}$ \\
		
		5.7659 & -0.2356 & 0.0018 \\
		
		6.0109 & -0.3359 & $4.7778\times10^{-5}$ \\
		
		6.2182 & -0.5196 & $9.6457\times10^{-4}$ \\
		
		6.3842 & -0.9159 & $1.5751\times10^{-5}$ \\
		
		6.5055 & -2.0524 & $2.8054\times10^{-4}$ \\
		
		6.5794 & -8.7586 & $1.6729\times10^{-6}$ \\
		
		\hline
		
	\end{tabular}
	
\end{table}

We want to stress here, that the preprint \cite{demiguel2026limitsaxiondarkmatter} does not contain enough information in order to fairly evaluate the presented results.
For being able to better evaluate and understand the performance of dielectric haloscopes, in general we recommend to carefully asses the following:

\begin{itemize}
\item The calculation of the coupling of the axion field to the mode being investigated should be transparently described. Uncertainties on this coupling need to be considered in sensitivity estimations and limit setting.
\item A qualitative confirmation of the expected field shape of the reflectivity induced mode used for limit setting should be performed, for example, using the bead pull method \cite{Egge_2024}. 
\item The Reflectivity of the setup should be measured and displayed in a broad frequency range where one can see other potential features (modes) besides the peak under consideration (as the reflectivity peak shown in Fig.\,3 of \cite{demiguel2026limitsaxiondarkmatter}).
\item The initial calibrated power spectrum should be displayed next to the normalized power excess (like shown in Fig. 4 of the preprint). Again, this measurement should be conducted over an extended frequency range\footnote{With the setup described in the preprint, due to the usage of an LNF LNA  which is usually not impedance matched to 50 Ohm, and without a circulator, Fabry-Perot oscillations of the baseline should be visible. These should be considered for the evaluation of the axion induced resonance \cite{MADMAX1}.} 

\end{itemize}

As a conclusion we state that based on the information available in the preprint of \cite{demiguel2026limitsaxiondarkmatter}, we are not able to reproduce the sensitivity of the described setup to ALP or hidden photon dark matter as given in Fig. 5 of \cite{demiguel2026limitsaxiondarkmatter} using the known and published formalisms that are available to us \cite{MADMAX1, MADMAX2, MADMAX3, Egge_2024, Egge_2023, MADMAX4, MADMAX_PoP_2020, Knirck_2019, Millar_2017, MADMAX_PRL_2017}.

{\small
\renewcommand{\refname}{}
\bibliographystyle{unsrt}
\bibliography{refs}
}


\end{document}